\documentclass[amsmath,notitlepage,amssymb,aps,showkeys,floatfix,prd,a4paper,
  twocolumn,nofootinbib]{revtex4-1}

\usepackage{graphicx}
\usepackage{amsmath}
\usepackage{epstopdf}
\usepackage{amsfonts,amssymb}
\usepackage[utf8]{inputenc}
\usepackage{pstricks}
\usepackage{color}

\usepackage[compat=1.0.0]{tikz-feynman}
\usepackage{simplewick}
\usepackage{color, colortbl}
\definecolor{Gray}{gray}{0.9}
\usepackage{float}
\usepackage{tikz-feynman}
\usepackage{feynmp}
\usepackage{forest}

\usepackage{verbatim}    
\usepackage{dcolumn}
\usepackage{bm}
\usepackage{epsfig}
\usepackage{braket}
\usepackage[normalem]{ulem} 
\usepackage{longtable} 

\newcommand{\be}{\begin{equation}}
\newcommand{\ee}{\end{equation}}
\newcommand{\ben}{\begin{eqnarray}}
\newcommand{\een}{\end{eqnarray}}

\def\MeV{\mbox{ MeV}}

\def\MeV{\mbox{ MeV}}

\newcommand{\pslash}{\not{\hbox{\kern-2.3pt $p$}}}
\newcommand{\pdslash}{\not{\hbox{\kern-2pt $\partial$}}}

\begin{document}

\title{The multiplicity of the doubly charmed state $T_{cc} ^+$ in heavy-ion
  collisions}

\author{ L. M. Abreu}
\email{luciano.abreu@ufba.br}
\affiliation{ Instituto de F\'isica, Universidade Federal da Bahia,
Campus Universit\'ario de Ondina, 40170-115, Bahia, Brazil}

\author{F. S. Navarra}
\email{navarra@if.usp.br}
\affiliation{Instituto de F\'{\i}sica, Universidade de S\~{a}o Paulo, 
Rua do Mat\~ao, 1371 \\ CEP 05508-090,  S\~{a}o Paulo, SP, Brazil}


\author{H. P. L. Vieira}
\email{hildeson.paulo@ufba.br}
\affiliation{ Instituto de F\'isica, Universidade Federal da Bahia,
Campus Universit\'ario de Ondina, 40170-115, Bahia, Brazil}

\begin{abstract}

  We study the evolution of the  doubly charmed state 
  $T_{cc}^+$ in a hot hadron gas produced in the late stage of heavy-ion
  collisions. We use effective Lagrangians to calculate the thermally 
  averaged cross sections of $T_{cc}^+$ production in reactions  such as
  $  D^{(*)} D^{(*)}  \rightarrow T_{cc}^+ \pi,  T_{cc}^+ \rho  $  and its
  absorption in the corresponding inverse processes.  We then solve the 
  rate equation to follow the time evolution of the $T_{cc}^+$ multiplicity,
  and determine how it is affected by the considered reactions during the
  expansion of the hadronic matter. We compare the evolution of the $T_{cc}^+$ 
  abundance treated as a hadronic $S$-wave molecule and as a tetraquark
  state.  Our results show that the tetraquark yield increases by a
  factor of about 2 at freeze-out, but it is still almost two orders of magnitude
  smaller than the final yield of molecules formed from hadron coalescence.     
  We also analyze the dependence of the yields with the system
  size,  represented by
  $\mathcal{N} = \left[ d N_{ch} / d \eta (\eta < 0.5)\right]^{1/3}$.
 We make predictions which can be confronted with data, when they
    are available. 

\end{abstract}

\maketitle

\section{Introduction}
\label{Introduction} 


The LHCb collaboration reported a few months ago the observation of the first 
doubly charmed tetraquark state in proton-proton $(pp)$ collisions, with
statistical significance of more than $10 \,
\sigma$~\cite{LHCb:2021vvq,LHCb:2021auc}. It has been identified from the fit
of a narrow peak seen in the $D^0 D^0 \pi^+ $-mass spectrum to one resonance
with a mass of approximately $3875 \MeV$ and quantum numbers $J^P = 1^+$.
Its minimum valence quark content is $c c \bar{u}\bar{d} $. From the data, the 
binding energy 
with respect to the $D^{*+} D^0 $ mass threshold and the decay width are
estimated to be
$273\pm 61 \pm 5 _{-14}^{+11}$ keV and          
$410\pm 165 \pm 43 _{-38}^{+18}$ keV, respectively, which are consistent with 
the expected properties of a $T_{cc}^+ $ isoscalar tetraquark ground state
with $J^P = 1^+$~\cite{Gelman:2002wf,Janc:2004qn,Vijande:2003ki,Navarra:2007yw,
Vijande:2007rf,Ebert:2007rn,Lee:2009rt,Yang:2009zzp,Hong:2018mpk,    
Hudspith:2020tdf,Cheng:2020wxa,Qin:2020zlg}. Since its detection, many works
appeared debating the possible mechanisms of its decay/formation and trying to
answer the question of whether it is an extended hadron molecule
or a compact tetraquark~\cite{Agaev:2021vur,Dong:2021bvy, 
Agaev:2021vur,Dong:2021bvy,Huang:2021urd,Li:2021zbw,Ren:2021dsi,Xin:2021wcr, 
Yang:2021zhe,Meng:2021jnw,Ling:2021bir,Fleming:2021wmk,Jin:2021cxj, 
Azizi:2021aib,Hu:2021gdg,Abreu:2021jwm,Albaladejo:2021vln,Dai:2021vgf}.

To reach a more complete picture of the $T_{cc}^+ $ state, additional
experimental and theoretical work is needed.  
In this sense, in Ref.~\cite{Abreu:2021jwm} we have suggested that a good
environment to study the $T_{cc}^+$ properties is the one provided by heavy
ion collisions (HIC),  where a large number of charm quarks is produced. 
In a typical HIC, there is a phase transition from nuclear matter to the  
locally thermalized state of deconfined quarks and gluons -- the so-called 
quark-gluon plasma (QGP), which expands, cools down and becomes a gas of   
hadrons. In this last transition heavy quarks coalesce to form multiquark
bound states. As the hadronic phase evolves, the multiquark states interact
with other hadrons, being destroyed in collisions with the comoving light
mesons, or produced through the inverse processes~
\cite{Chen:2007zp,ChoLee1,XProd1,XProd2,UFBaUSP1,MartinezTorres:2017eio,   
  Abreu:2017cof,Abreu:2018mnc,LeRoux:2021adw}. At this stage, the spatial    
configuration of the multiquark systems influences the hadronic interactions
and therefore the final yields. 
More concretely: similarly to the $X(3872)$ state (see the discussion 
in Ref.~\cite{XProd2}), charm meson molecules $D D^*$ are larger than charm    
tetraquarks in a diquark-antidiquark configuration $(c c) -  (\bar{q}\bar{q})$ 
by a factor about 3-10, and therefore their absorption cross sections may be  
one order of magnitude larger. In contrast, when the $T_{cc} $ is produced from 
$D - D^*$ fusion in a hadron gas,  the initial $D \, + \,D^*$ state has a bigger
spatial overlap with a molecule than with a tetraquark. For this reason,   
molecules are expected to be more easily produced as well as more easily
destroyed than compact tetraquarks in a hot hadronic environment.

To the best of our knowledge, the interactions of the $T_{cc}$ in a hadronic
medium  were first discussed in Ref.~\cite{Hong:2018mpk} (published before the
observation of the $T_{cc}$  by the LHCb Collaboration). The $T_{cc}$ was
treated as an extremely shallow bound state of a $D$ and a $D^*$.  
The authors used the quasi-free approximation, in which the charm mesons are
understood to be on-shell and their binding energy and mutual interaction are
neglected. In this
approximation, the $T_{cc}$ is absorbed when a pion from the hadron gas
interacts with the $D$ or with the $D^*$. In this approach , the dynamical 
component needed is just the effective $D^* \, D \, \pi$ Lagrangian. The    
results suggested that the hadronic effects on the $T_{cc}$ final abundance 
depend on the initial yield of $T_{cc}$  produced from the quark-gluon plasma
phase, which is determined by the assumed structure of the state.

We believe that the  subject deserves further discussion.  
In a recent paper~\cite{Abreu:2021jwm}, we computed the cross sections of 
$T_{cc}$ production in reactions such as 
$  D^{(*)} D^{(*)}  \rightarrow T_{cc} \pi,  T_{cc} \rho  $  and its    
absorption in the corresponding inverse processes. The absorption cross
sections were found to be larger than the production ones.   
These results were obtained  using effective field Lagrangians to account
for the couplings between light and heavy mesons. For the new state there is
no Lagrangian and we had to study the  $T_{cc}-D-D^*$ with QCD sum rules,
determining, for the first time,  the form factor and the coupling constant.

In Ref.~\cite{Abreu:2021jwm} the time evolution of the $T_{cc}$ abundance in
the hot hadron gas  was not addressed.  Thus, in this work we complete the
work done in Ref.~\cite{Abreu:2021jwm}. 
We calculate the thermally averaged cross sections of $T_{cc}^+$ production
and absorption, and use them as input to solve the kinetic equation and obtain 
the time evolution of the $T_{cc}^+$ multiplicity. We compute  
the $T_{cc}^+$  abundance considering it as a hadronic molecule and also as a
tetraquark state and compare them. Also, we present a comparison between the
time evolution of the multiplicities of $T_{cc} $ and  of $ X(3872)$ in similar
conditions. We finish with a
discussion on the dependence of the ratio $R= N_{T_{cc}}  / N_{X(3872)} $  
with the multiplicity density of charged particles measured at midrapidity
$\mathcal{N} = \left[ d N_{ch} / d \eta (\eta < 0.5)\right]^{1/3}$.

The paper is organized as follows. In Section~\ref{ThermalAvCrossSection} we 
discuss the cross sections averaged over the thermal distributions. In      
Section~\ref{abundance} we investigate the time evolution of the $X(3872)$
abundance  by  solving  the  kinetic  equation. Finally,                 
Section~\ref{Conclusions} is devoted to the summary and to the
concluding remarks.


\section{Cross sections averaged over the thermal distributions}

\label{ThermalAvCrossSection} 


\begin{figure}[!ht]
    \centering

\begin{tikzpicture}
\begin{feynman}
\vertex (a1) {$T_{cc} (p_{1})$};
	\vertex[right=1.5cm of a1] (a2);
	\vertex[right=1.cm of a2] (a3) {$D (p_{3})$};
	\vertex[right=1.4cm of a3] (a4) {$T_{cc} (p_{1})$};
	\vertex[right=1.5cm of a4] (a5);
	\vertex[right=1.cm of a5] (a6) {$D (p_{3})$};
\vertex[below=1.5cm of a1] (c1) {$\pi (p_{2})$};
\vertex[below=1.5cm of a2] (c2);
\vertex[below=1.5cm of a3] (c3) {$D (p_{4})$};
\vertex[below=1.5cm of a4] (c4) {$\pi (p_{2})$};
\vertex[below=1.5cm of a5] (c5);
\vertex[below=1.5cm of a6] (c6) {$D (p_{4})$};
	\vertex[below=2cm of a2] (d2) {(a)};
	\vertex[below=2cm of a5] (d5) {(b)};
\diagram* {
  (a1) -- (a2), (a2) -- (a3), (c1) -- (c2), (c2) -- (c3), (a2) --
  [fermion, edge label'= $D^{*}$] (c2), (a4) -- (a5), (a5) -- (c6),
  (c4) -- (c5), (c5) -- (a6), (a5) -- [fermion, edge label'= $D^{*}$] (c5)
}; 
\end{feynman}
\end{tikzpicture}

\begin{tikzpicture}
\begin{feynman}
\vertex (a1) {$T_{cc} (p_{1})$};
	\vertex[right=1.5cm of a1] (a2);
	\vertex[right=1.cm of a2] (a3) {$D^* (p_{3})$};
	\vertex[right=1.4cm of a3] (a4) {$T_{cc}  (p_{1})$};
	\vertex[right=1.5cm of a4] (a5);
	\vertex[right=1.cm of a5] (a6) {$D^* (p_{3})$};
\vertex[below=1.5cm of a1] (c1) {$\pi (p_{2})$};
\vertex[below=1.5cm of a2] (c2);
\vertex[below=1.5cm of a3] (c3) {$D^* (p_{4})$};
\vertex[below=1.5cm of a4] (c4) {$\pi (p_{2})$};
\vertex[below=1.5cm of a5] (c5);
\vertex[below=1.5cm of a6] (c6) {$D^* (p_{4})$};
	\vertex[below=2cm of a2] (d2) {(c)};
	\vertex[below=2cm of a5] (d5) {(d)};
\diagram* {
  (a1) -- (a2), (a2) -- (a3), (c1) -- (c2), (c2) -- (c3), (a2) --
  [fermion, edge label'= $D$] (c2), (a4) -- (a5), (a5) -- (c6), (c4) --
  (c5), (c5) -- (a6), (a5) -- [fermion, edge label'= $D$] (c5)
}; 
\end{feynman}
\end{tikzpicture}

\begin{tikzpicture}
\begin{feynman}
\vertex (a1) {$T_{cc} (p_{1})$};
	\vertex[right=1.5cm of a1] (a2);
	\vertex[right=1.cm of a2] (a3) {$D^* (p_{3})$};
	\vertex[right=1.4cm of a3] (a4) {$T_{cc} (p_{1})$};
	\vertex[right=1.5cm of a4] (a5);
	\vertex[right=1.cm of a5] (a6) {$D^* (p_{3})$};
\vertex[below=1.5cm of a1] (c1) {$\rho (p_{2})$};
\vertex[below=1.5cm of a2] (c2);
\vertex[below=1.5cm of a3] (c3) {$D (p_{4})$};
\vertex[below=1.5cm of a4] (c4) {$\rho (p_{2})$};
\vertex[below=1.5cm of a5] (c5);
\vertex[below=1.5cm of a6] (c6) {$D (p_{4})$};
	\vertex[below=2cm of a2] (d2) {(e)};
	\vertex[below=2cm of a5] (d5) {(f)};
\diagram* {
  (a1) -- (a2), (a2) -- (a3), (c1) -- (c2), (c2) -- (c3), (a2) --
  [fermion, edge label'= $D$] (c2), (a4) -- (a5), (a5) -- (c6), (c4) --
  (c5), (c5) -- (a6), (a5) -- [fermion, edge label'= $D^{*}$] (c5)
}; 
\end{feynman}
\end{tikzpicture}

\begin{tikzpicture}
\begin{feynman}
\vertex (a1) {$T_{cc} (p_{1})$};
	\vertex[right=1.5cm of a1] (a2);
	\vertex[right=1.cm of a2] (a3) {$D (p_{3})$};
	\vertex[right=1.4cm of a3] (a4) {$T_{cc} (p_{1})$};
	\vertex[right=1.5cm of a4] (a5);
	\vertex[right=1.cm of a5] (a6) {$D (p_{3})$};
\vertex[below=1.5cm of a1] (c1) {$\pi (p_{2})$};
\vertex[below=1.5cm of a2] (c2);
\vertex[below=1.5cm of a3] (c3) {$D^* (p_{4})$};
\vertex[below=1.5cm of a4] (c4) {$\rho (p_{2})$};
\vertex[below=1.5cm of a5] (c5);
\vertex[below=1.5cm of a6] (c6) {$D (p_{4})$};
	\vertex[below=2cm of a2] (d2) {(g)};
	\vertex[below=2cm of a5] (d5) {(h)};
\diagram* {
  (a1) -- (a2), (a2) -- (a3), (c1) -- (c2), (c2) -- (c3), (a2) --
  [fermion, edge label'= $D^*$] (c2),(a4) -- (a5), (a5) -- (a6),
  (c4) -- (c5), (c5) -- (c6), (a5) -- [fermion, edge label'= $D^*$] (c5)
}; 
\end{feynman}
\end{tikzpicture}

\begin{tikzpicture}
\begin{feynman}
\vertex (a1) {$T_{cc} (p_{1})$};
	\vertex[right=1.5cm of a1] (a2);
	\vertex[right=1.cm of a2] (a3) {$D (p_{3})$};
	\vertex[right=1.4cm of a3] (a4) {$T_{cc} (p_{1})$};
	\vertex[right=1.5cm of a4] (a5);
	\vertex[right=1.cm of a5] (a6) {$D^{*} (p_{3})$};
\vertex[below=1.5cm of a1] (c1) {$\rho (p_{2})$};
\vertex[below=1.5cm of a2] (c2);
\vertex[below=1.5cm of a3] (c3) {$D (p_{4})$};
\vertex[below=1.5cm of a4] (c4) {$\rho (p_{2})$};
\vertex[below=1.5cm of a5] (c5);
\vertex[below=1.5cm of a6] (c6) {$D^{*} (p_{4})$};
	\vertex[below=2cm of a2] (d2) {(i)};
	\vertex[below=2cm of a5] (d5) {(j)};

\diagram* {
  (a1) -- (a2), (a2) -- (c3), (c1) -- (c2), (c2) -- (a3), (a2) -- 
  [fermion, edge label'= $D^*$] (c2), (a4) -- (a5), (a5) -- (a6),
  (c4) -- (c5), (c5) -- (c6), (a5) -- [fermion, edge label'= $D$] (c5)
}; 
\end{feynman}
\end{tikzpicture}

\begin{tikzpicture}
\begin{feynman}
\vertex (a1) {$T_{cc} (p_{1})$};
	\vertex[right=1.5cm of a1] (a2);
	\vertex[right=1.cm of a2] (a3) {$D^* (p_{3})$};
\vertex[below=1.5cm of a1] (c1) {$\rho (p_{2})$};
\vertex[below=1.5cm of a2] (c2);
\vertex[below=1.5cm of a3] (c3) {$D^* (p_{4})$};
	\vertex[below=2cm of a2] (d2) {(k)};
\diagram* {
  (a1) -- (a2), (a2) -- (c3), (c1) -- (c2), (c2) -- (a3), (a2) --
  [fermion, edge label'= $D$] (c2)
}; 
\end{feynman}
\end{tikzpicture}
\caption{Reproduction of Born diagrams treated in                 
  Ref.~\cite{Abreu:2021jwm} contributing to the following process
  (without specification of the charges of the particles):     
  $ T_{cc} \pi \rightarrow D D $  [(a) and (b)],     
  $ T_{cc} \pi \rightarrow D^* D^* $  [(c) and (d)],
  $ T_{cc} \rho \rightarrow D^* D $  [(e) and (f)],
  $ T_{cc} \pi \rightarrow D D^* $  [(g)],           
  $ T_{cc} \rho \rightarrow D D $  [(h) and (i)] and  
  $ T_{cc} \rho \rightarrow D^* D^* $  [(j) and (k)].
  The particle charges are not specified. }
\label{DIAG1}
\end{figure}
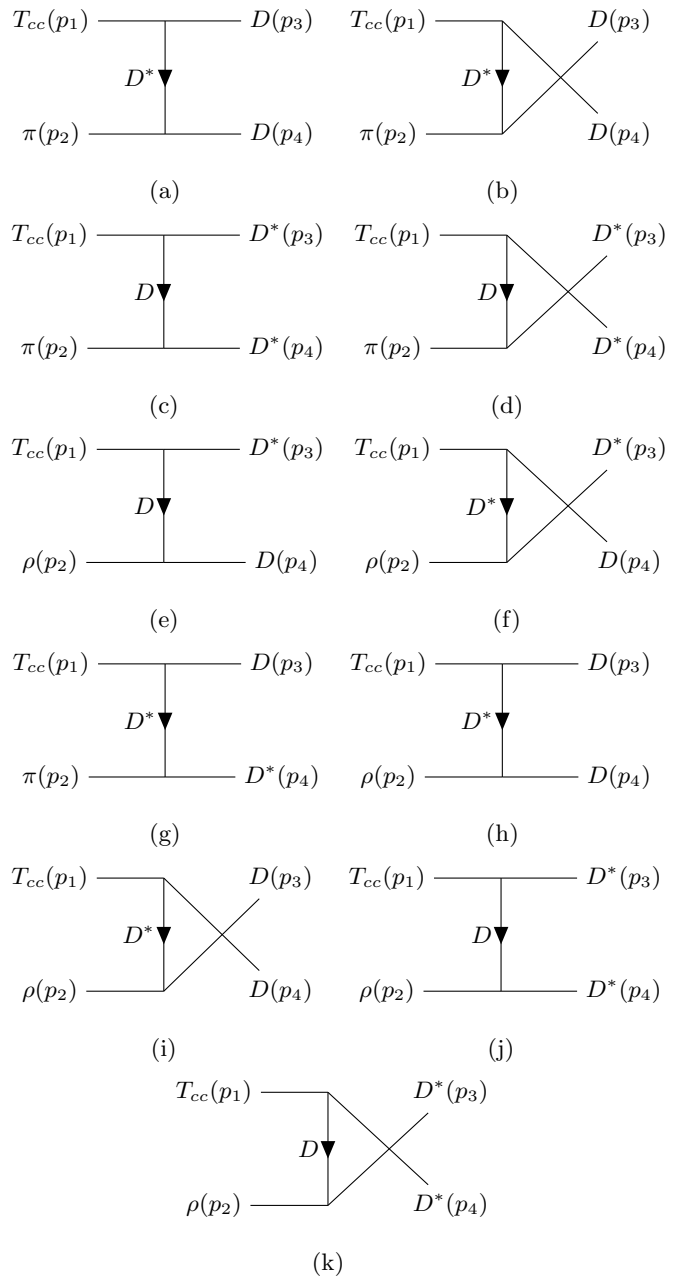



We are interested in the hadronic effects on the $T_{cc}^+$ state in a  
heavy-ion collision environment, in particular in how its multiplicity      
is affected by the production and absorption reactions during the expansion 
of the hadron gas. The interactions of $T_{cc}^+$ with
other hadrons are described with the help of the results reported in our previous 
work~\cite{Abreu:2021jwm}, especially focusing on the reactions 
$ T_{cc}^+ \pi \rightarrow D^{(*)} D^{(*)} $ and                             
$ T_{cc}^+ \rho \rightarrow D^{(*)} D^{(*)} $, as well as the corresponding
inverse processes. We reproduce the lowest-order Born diagrams considered 
in Fig.~\ref{DIAG1}. In the diagrams $(a)-(f)$, the vertices involving light
and heavy-light mesons are described by effective Lagrangians of the type  
$\mathcal{L}_{PPV}$ and $\mathcal{L}_{VVV}$,  where $P$ and $V$ are        
pseudoscalar and  vector mesons, respectively. In the case of the
diagrams $(g)-(k)$, the vertices involving light and heavy-light mesons are 
anomalous, i.e. they must be of the type
$\mathcal{L}_{PVV}$~\cite{Abreu:2021jwm}. 
Taking into consideration that the  $T_{cc}^+$ has quantum numbers
$I (J^P) = 0 (1^+) $, its effective coupling  with the $D D^*$ pair is
of the form
$\mathcal{L}_{T_{cc}} = ig_{T_{cc} D D^*} T_{cc}^{\mu}  D_{\mu}^{*} D $
~\cite{Ling:2021bir,Abreu:2021jwm}, where 
 $T_{cc}$ denotes the field associated to  
$T_{cc}^+$ state; this notation will be used in what follows. Also, the       
$D_{\mu}^{*}  D $ represents the $D_{\mu}^{*+} D^0 $ and  $D_{\mu}^{*0} D^+ $ 
components, although we do not distinguish them here since we will use
isospin-averaged masses. As mentioned above,  in~\cite{Abreu:2021jwm} we
have determined the form factor and the corresponding coupling constant
associated to the $T_{cc}-D-D^*$ vertex with QCD sum rules (QCDSR).
A comprehensive study on the $T_{cc}$ cross sections has been performed.

All the aforementioned reactions happen in a hadron gas at finite temperature,
which should drive the collision energies of the colliding particles. As a
consequence, the relevant dynamical quantity is the cross section averaged over
the thermal distribution for a reaction involving an initial two-particle state
going into two final particles $ab \to cd$. It is defined
as~\cite{Koch,ChoLee1,XProd2}
\begin{eqnarray}
  \langle \sigma_{a b \rightarrow c d } \,  v_{a b}\rangle &  = & \frac{ \int 
    d^{3} \mathbf{p}_a  d^{3} \mathbf{p}_b \, f_a(\mathbf{p}_a) \,
    f_b(\mathbf{p}_b) \, 
    \sigma_{a b \rightarrow c d } \,\,v_{a b} }{ \int  d^{3} \mathbf{p}_a  d^{3}
    \mathbf{p}_b \, f_a(\mathbf{p}_a) \,  f_b(\mathbf{p}_b) }
\nonumber \\
& = & \frac{1}{4 \alpha_a ^2 K_{2}(\alpha_a) \alpha_b ^2 K_{2}(\alpha_b) }
\int _{z_0} ^{\infty } dz  K_{1}(z) \,\,
\nonumber \\
& & \times \sigma (s=z^2 T^2) \left[ z^2 - (\alpha_a + \alpha_b)^2 \right]
\nonumber \\
& &\left[ z^2 - (\alpha_a - \alpha_b)^2 \right],
\label{ThermalCrossSection}
\end{eqnarray}
where $v_{ab}$ represents the relative velocity of the two initial  interacting 
particles $a$ and $b$; $\sigma_{ab \to cd}$  denotes the cross sections   
evaluated in \cite{Abreu:2021jwm} for the different reactions shown in
Fig.~\ref{DIAG1}; the function $f_i(\mathbf{p}_i)$ is the Bose-Einstein   
distribution of particles of species $i$, which depends on the temperature    
$T$; $\beta _i = m_i / T$, $z_0 = max(\beta_a + \beta_b,\beta_c 
+ \beta_d)$; and  $K_1$ and $K_2$ the modified Bessel functions.

\begin{figure}[!ht]
    \centering
       \includegraphics[{width=1.0\linewidth}]{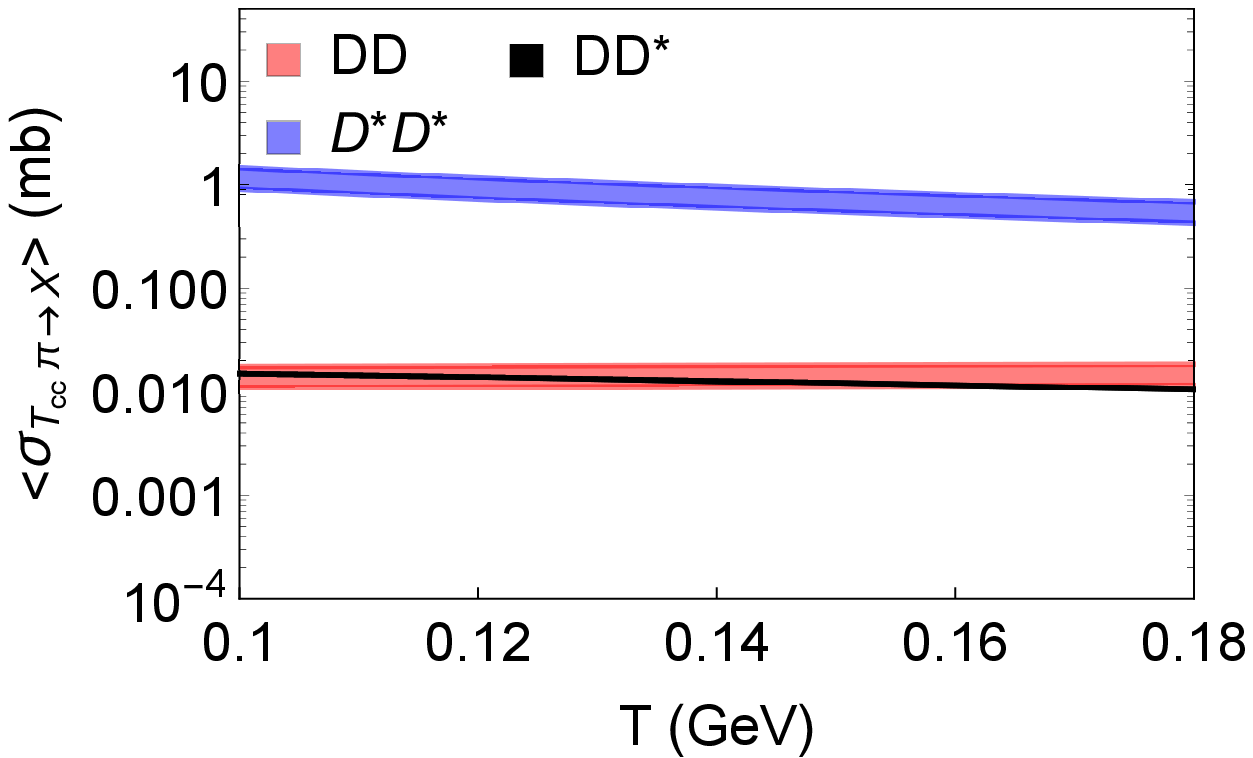} 
       \includegraphics[{width=1.0\linewidth}]{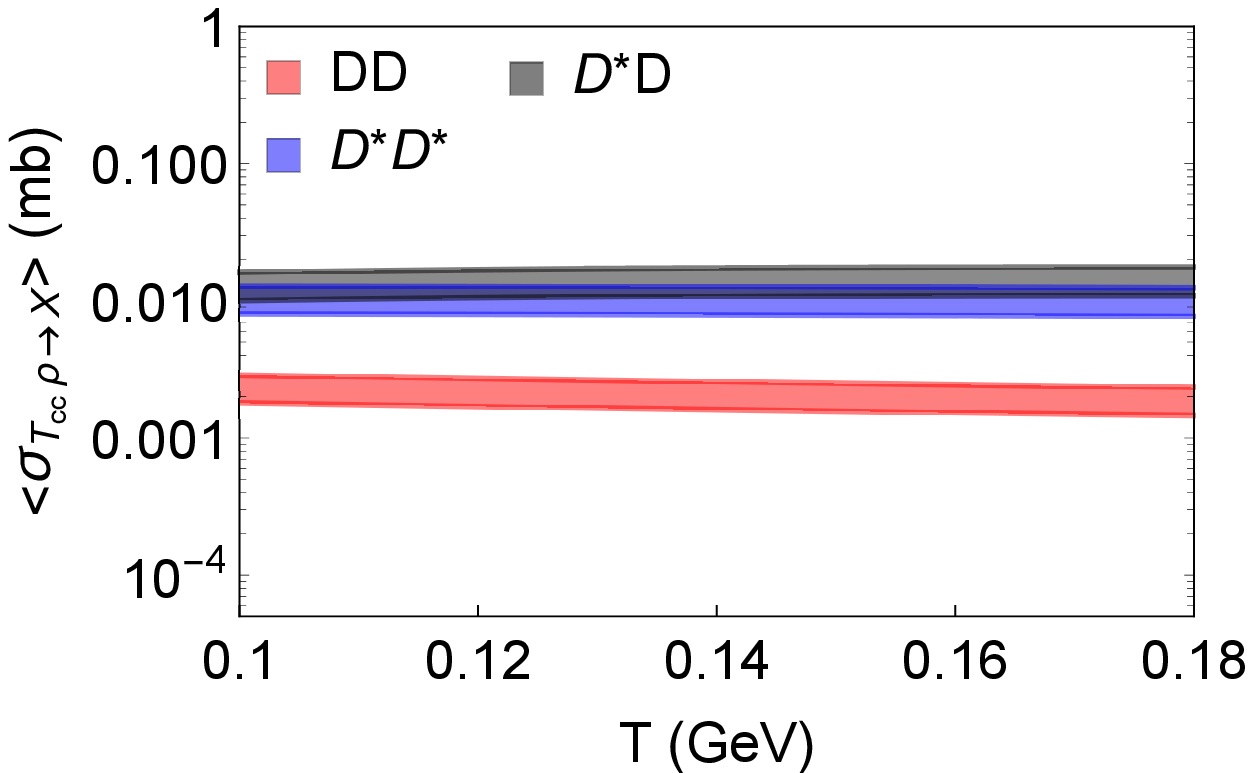} 
    \caption{Thermally averaged cross sections for the absorption processes $T_{cc}^+ \pi \rightarrow D^{(*)} D^{(*)} $ (top panel) and  $T_{cc}^+ \rho \rightarrow D^{(*)} D^{(*)} $ (bottom panel), as a function of temperature $T$. Upper and lower limits of the bands are obtained taking the upper and lower limits of the uncertainty in the coupling constant $g_{T_{cc}DD^*}$~\cite{Abreu:2021jwm}. In the top panel, the band for the $DD$ and $D D^*$ channels coincide.  }
    \label{Fig:AvCrSec-Abs}
\end{figure}

\begin{figure}[!ht]
    \centering
       \includegraphics[{width=1.0\linewidth}]{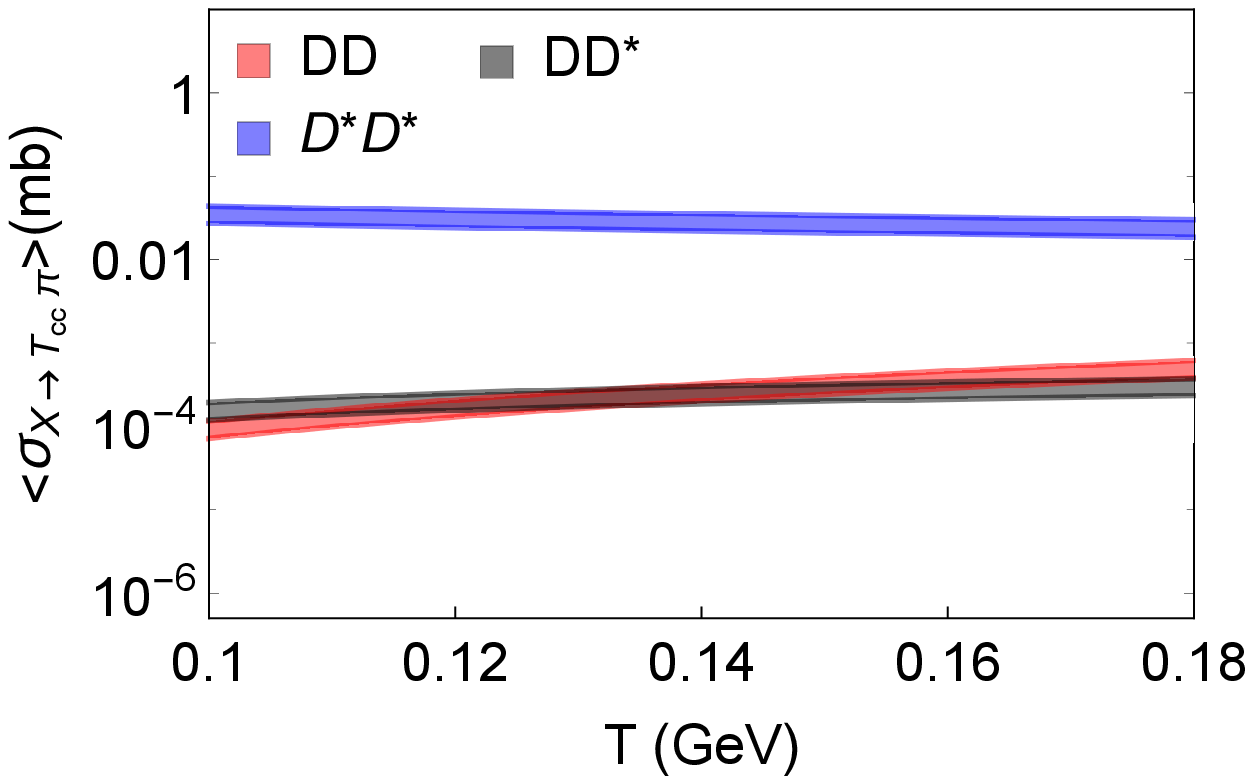}
       \includegraphics[{width=1.0\linewidth}]{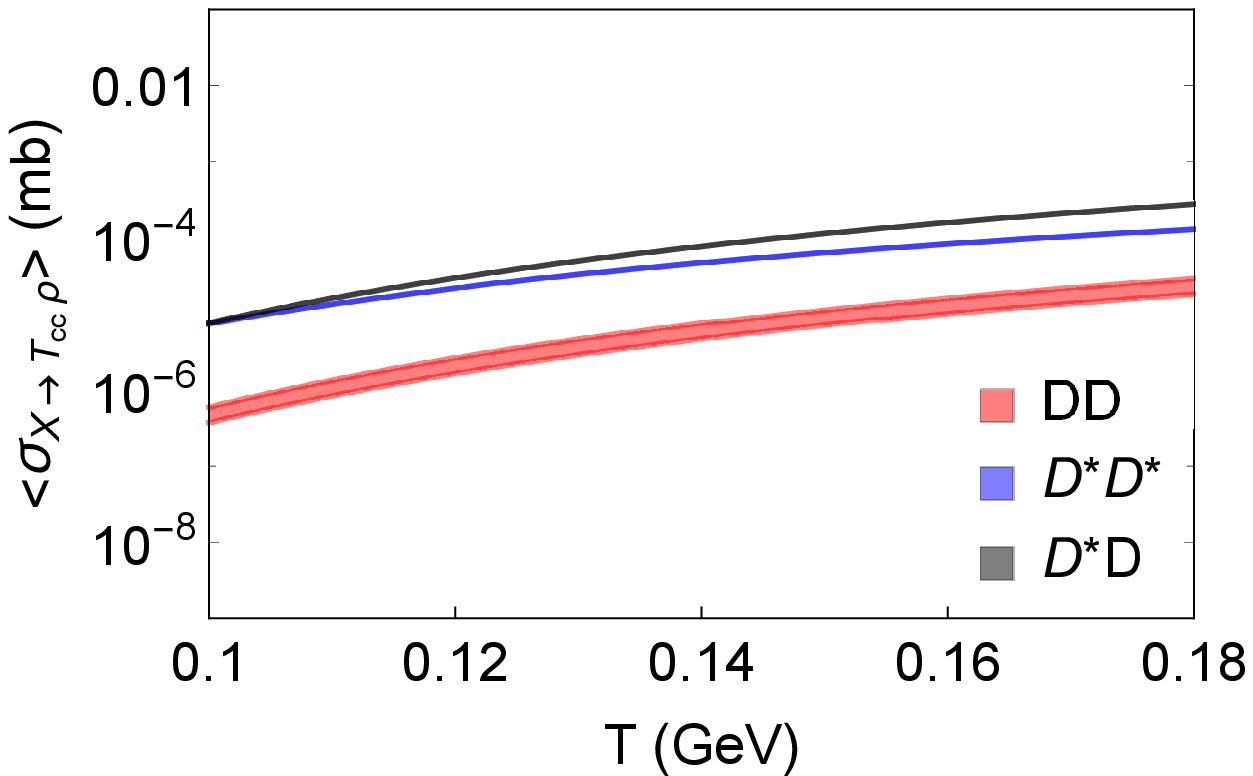}
       \caption{Thermally averaged cross sections as a function of temperature 
         for the respective inverse (production) processes displayed in 
         Fig.~\ref{DIAG1}, i.e.
         $D^{(*)} D^{(*)}\rightarrow T_{cc}^+ \pi  $ (top panel) and
         $D^{(*)} D^{(*)} \rightarrow T_{cc}^+ \rho $  (bottom panel),      
         obtained via the detailed balance relation. Upper and lower limits 
         of the bands are obtained taking the upper and lower limits of the 
         uncertainty in the coupling constant
         $g_{T_{cc}DD^*}$~\cite{Abreu:2021jwm}.} 
    \label{Fig:AvCrSec-Prod}
\end{figure}

In Figs.~\ref{Fig:AvCrSec-Abs} and ~\ref{Fig:AvCrSec-Prod} we plot the 
thermally averaged cross sections as functions of the temperature for        
$T_{cc}$ absorption and production respectively, via the processes discussed
above. The bands in the figures express the uncertainty in the coupling constant 
$g_{T_{cc}DD^*}$ coming from variations in the quantities relevant in the
QCDSR calculations ~\cite{Abreu:2021jwm}. 
It can be seen that in general the reactions involving a pion in initial    
or final state have greater thermal cross sections than those with a $\rho$
meson. Interestingly, the thermal cross sections
$\langle \sigma_{a b \rightarrow c d } \,  v_{a b}\rangle$ for $T_{cc}$ 
absorption  do not change appreciably in the considered range
of temperature (remaining almost constant), when compared to corresponding
ones for the $T_{cc}$ production. 
Also, the results suggest that the channels $T_{cc} \pi \rightarrow  D D,
D D^{*}$ have similar magnitudes, with the final state $D^*D^*$ being enhanced
with respect to these other ones by almost two orders of magnitude. 
On the other hand, considering the uncertainty, the channels for $T_{cc} \rho    
\rightarrow  D D^*, D^* D^{*}$ have similar magnitudes, whereas the final
state $T_{cc} \rho \rightarrow  D D$ is smaller by almost one order of magnitude. 

The most important conclusion from these figures is that the thermally      
averaged cross sections for  $T_{cc}$ annihilation  are bigger   
than those for production, at least by  one or two orders of magnitude.
This might play an important role in the time evolution of the  $T_{cc}$ 
multiplicity. Thus, in the next section we use these thermally averaged    
cross sections as input in the rate equation and study the time evolution
of the doubly charmed state abundance.

\begin{figure}[!ht]
    \centering
       \includegraphics[{width=1.0\linewidth}]{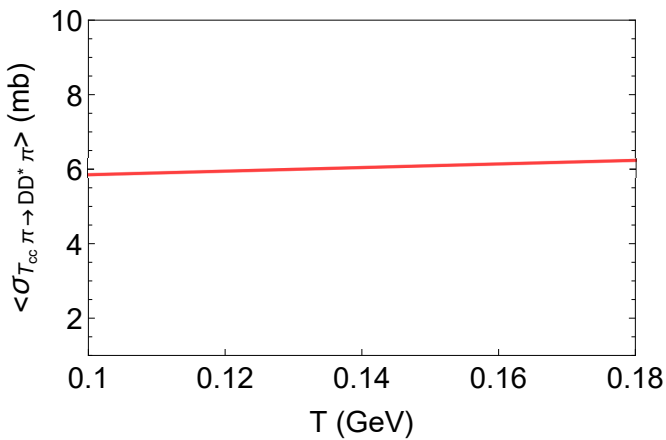}
\caption{ Thermal cross sections as a function of the temperature for the
$T_{cc}$-absorption by pions in the quasi-free approximation, which, 
according to Eq.~(26) of Ref.~\cite{Hong:2018mpk}  is given by
$\langle \sigma_{T_{cc} \pi \rightarrow D D^{*} \pi} \,  v_{T_{cc} \pi}\rangle 
= \langle \sigma_{D \pi \rightarrow D \pi} \,  v_{T_{cc} \pi}\rangle  +
\langle \sigma_{D^* \pi \rightarrow D^* \pi} \,  v_{T_{cc} \pi}\rangle $ .}
    \label{Fig:AvCrSec-Hong}
\end{figure}

We would like to close this  section with a comparison between
our findings and those reported in Ref.~\cite{Hong:2018mpk}.
In the approach developed on ~\cite{Hong:2018mpk}, the $T_{cc}$ is absorbed
when a pion from the hadron gas interacts either with the $D$ or with the $D^*$.
In each of these interactions the other heavy meson is a spectator. As discussed
in Ref.~\cite{Abreu:2021jwm}, this approach has the advantage of relying solely 
on the well-known $D^* \, D \, \pi$ Lagrangian. However it ignores the possible
dynamical effects associated with  the quantum numbers of the $D^* \, D$ bound 
state. Moreover it does not include some possible final states. We remember that, 
as pointed out in Ref.~\cite{Abreu:2021jwm}, the cross sections for the
absorption of the $T_{cc}$ by pions in the quasi-free approximation are much
larger than those obtained with the present approach. For completeness, we show
in Fig.~\ref{Fig:AvCrSec-Hong}  the thermal  cross section for the
$T_{cc}$-absorption by pions in the quasi-free approximation. As expected, it is 
much larger than the  one shown in Fig.~\ref{Fig:AvCrSec-Abs}.


\section{Time evolution of the $T_{cc}^+$ abundance }

\label{abundance}




\subsection{The rate equation} 

Now we study the effect of the  $\pi (\rho) - T_{cc}$  interactions on the
abundance of $T_{cc}^+$ during the hadron gas phase of heavy ion collisions.   
The momentum-integrated evolution equation for the $T_{cc}$ abundance
reads~\cite{ChoLee1,XProd2,Koch}  
\begin{eqnarray} 
  \frac{ d N_{T_{cc}} (\tau)}{d \tau} & = & \sum_{\substack{c, c^{\prime} = D,
      D^* \\ \varphi = \pi, \rho}} 
\left[ \langle \sigma_{c c^{\prime} \rightarrow T_{cc} \varphi } 
v_{c c^{\prime}} \rangle n_{c} (\tau) N_{c^{\prime}}(\tau)
 \right. \nonumber \\ & &  \left. 
 - \langle \sigma_{ \varphi T_{cc} \rightarrow c c^{\prime} }
 v_{ T_{cc}\varphi } 
\rangle n_{\varphi} (\tau) N_{T_{cc}}(\tau) 
\right], 
\label{rateeq}
\end{eqnarray}
where $N_{T_{cc}} (\tau)$, $N_{c^{\prime}}(\tau)$,  $ n_{c} (\tau)$ and 
$n_{\varphi} (\tau)$ are the abundances of $T_{cc}$ and of charmed            
mesons of type $c^{\prime}$, and the densities of  charmed mesons of type $c$
and of light mesons at proper time $\tau$, respectively.
Eq.~(\ref{rateeq}) implies that the time evolution of $N_{T_{cc}} (\tau)$      
depends on both the $T_{cc} $ dissociation and production rates  through the
processes  discussed previously.

To solve Eq.~(\ref{rateeq}) we assume that  the pions, $\rho$ and charmed 
mesons in the reactions contributing to the abundance of $T_{cc}$ are   
in equilibrium. Accordingly, $ n_{c} (\tau)$, $N_{c^{\prime}}(\tau)$ 
and $n_{\varphi} (\tau)$ can be written as~\cite{ChoLee1,XProd2,Koch}
\ben n_{i} (\tau) &  \approx & \frac{1}{2 \pi^2}\gamma_{i} g_{i} m_{i}^2 
T(\tau)K_{2}\left(\frac{m_{i} }{T(\tau)}\right), 
\label{densities}
\een
where $\gamma _i$ and $g_i$ are the fugacity factor, the degeneracy  
factor and $m_i$ the mass  of the particle  $i$, respectively. The multiplicity     
$N_i (\tau)$ is obtained by multiplying the density $n_i(\tau)$ by the volume
$V(\tau)$.

The time dependence of the density $n_{i} (\tau)$ is encoded in the
parametrization of the temperature $T(\tau)$ and of the volume $V(\tau)$,
which are fitted to reproduce the properties of the hadron gas.   According
to the boost invariant Bjorken picture, the hydrodynamical expansion and
cooling of the hadron gas is modeled as an accelerated transverse
expansion~\cite{ChoLee1,XProd2,Koch}, by the expressions (for $\tau \geq \tau _H $)
\ben
V(\tau) & = & \pi \left[ R_C + v_C \left(\tau - \tau_C\right) + 
\frac{a_C}{2} \left(\tau - \tau_C\right)^2 \right]^2 \tau_C , \nonumber \\
T(\tau) & = & T_C - \left( T_H - T_F \right) \left( \frac{\tau - 
\tau _H }{\tau _F - \tau _H}\right)^{\frac{4}{5}} .
\label{TempVol}
\een
where $R_C $ and $\tau_C$  denote the final transverse  and longitudinal   
sizes of the QGP; $v_C $ and  $a_C $ are its transverse flow velocity and 
transverse  acceleration at $\tau_C $; $T_C$ is the critical temperature of
the quark-hadron phase transition; $T_H $  is the temperature of the 
hadronic matter at the end of the mixed phase, occurring at the time 
$\tau_H $; and the kinetic freeze-out occurs at  $\tau _F $, when the
temperature
is $T_F $. We emphasize that this parametrization is employed as a proxy for 
capturing the basic elements of hydrodynamic expansion and cooling of the 
hadron matter, being adequate for our phenomenological approach, keeping in
mind that our focus is on the behavior of the $T_{cc}$ multiplicity during
the evolution of hadron gas phase. For a discussion of the features and
limitations of this model, we refer the reader to Ref.~\cite{Abreu:2017pos}. 
A more realistic hydrodynamical simulation is postponed to subsequent works.
\begin{center}
\begin{table}[h!]
\caption{Set of parameters used in Eq.~(\ref{TempVol}) for the hydrodynamic
expansion and cooling of the hadronic medium formed in  central
$Pb-Pb$ collisions at $\sqrt{s_{NN}} = 5.02$ TeV
~\cite{Abreu:2020ony,ExHIC:2017smd}.}
\vskip1.5mm
\label{param}
\begin{tabular}{ c c c }
\hline
\hline
 $v_C$ (c) & $a_C$ (c$^2$/fm) & $R_C$ (fm)   \\   
0.5 & 0.09 & 11  
\\  
\hline
 $\tau_C$ (fm/c) & $\tau_H$ (fm/c)  &  $\tau_F$ (fm/c)  \\   
7.1  & 10.2 & 21.5
\\  
\hline
  $T_C (\MeV)$  & $T_H (\MeV)$ & $T_F (\MeV)$ \\   
 156 & 156 & 115   \\  
\hline
 $N_c$  & $N_{\pi}(\tau_F)$ & $N_{\rho}(\tau_H)$ \\   
 14 & 2410 & 184 \\  
 \hline
 $V_C $ (fm${}^3$)  & &  \\   
   5380  &  &    \\  
\hline
\hline
\end{tabular}
\end{table}
\end{center}

\subsection{The initial conditions}

Unless explicitly stated otherwise, all the results refer to the
hadronic medium produced in central $Pb-Pb$ collisions at
$\sqrt{s_{NN}} = 5.02$ TeV
at the LHC.  We use in Eq. (\ref{TempVol})
the set of parameters of Ref.~\cite{Abreu:2020ony}, which has been
obtained in order to reproduce the quantities listed in Table 3.1 of
Ref.~\cite{ExHIC:2017smd}. The values of these parameters are given in
Table~\ref{param}.  We assume that the total number of charm quarks ($N_c$) 
in charmed hadrons is conserved during the production and dissociation    
reactions, i. e. $n_c(\tau) \times V(\tau) = N_c = const$. By doing this,
the charm quark fugacity factor $\gamma _c $ in Eq.~(\ref{densities}) is    
assumed to be time-dependent. In the case of pions and $\rho$ mesons, their
fugacities appear as normalization parameters, adjusted to fit the
multiplicities given in  Table~\ref{param}. 
We consider the yields obtained for the $T_{cc}$  with the coalescence model. 
In this model  the yield of a hadronic state depends on the overlap
of the density matrix of its constituents with its Wigner function.            
Consequently, this model encodes essential features of the internal structure,
such as angular momentum, number of constituent quarks, etc.         
Accordingly, the $T_{cc}$ multiplicity at the end of the quark-gluon plasma
phase is given by~\cite{ChoLee1,XProd2,Abreu:2020ony,ExHIC:2017smd}:
\ben
N_{T_{cc}} ^{Coal} & \approx & g_{T_{cc}} \prod _{j=1} ^{n} \frac{N_j}{g_j} 
\prod  _{i=1} ^{n-1} 
\frac{(4 \pi \sigma_i ^2)^{\frac{3}{2}} }{V (1 + 2 \mu _i T \sigma _i ^2 )} 
\nonumber \\
& & \times
\left[ \frac{4 \mu_i T \sigma_i ^2 }{3 (1 + 2 \mu _i T \sigma _i ^2 ) }
\right]^{l_i}, 
\label{TccCoal}
\een
where $g_j$ and $N_j$ are the degeneracy and number of the $j$-th constituent
of the $T_{cc}$ and 
$\sigma _i = (\mu _i \omega)^{-1/2}$. The quantity $\omega $ is the 
oscillator frequency (taking an harmonic oscillator as a picture for the hadron
internal structure) and $\mu$ the reduced mass, i.e.  
$\mu ^{-1} = m_{i+1} ^{-1}+ \left(\sum_{j=1} ^{i} m_j \right)^{-1}$. 
The angular momentum of the system, $l_i$, is 0 for an $S$-wave,  
and 1 for a $P$-wave. According to the coalescence model, the $T_{cc}$ is
produced 
as a S-wave tetraquark produced  at the end 
of the QGP phase at the critical temperature, when the volume is $V_C$. 
The oscillator frequency for tetraquark states        
produced via quark coalescence mechanism and the quark masses have been taken
to be $\omega_c = 220 \MeV$ and $m_q = 350 \MeV, m_c = 1500 \MeV$,
respectively~\cite{ExHIC:2017smd}. For molecular states, 
to calculate the oscillation frequency we have employed the expression
$\omega = 6 B$, with $B$ being the binding energy.
In Table II we give the multiplicities. 
For the sake of comparison, we have also included the multiplicities calculated
for the state $X(3872)$. 
We  have used in Eq.~(\ref{TccCoal}) the fact that if there are $N_c$ charm and $N_c$ anti-charm quarks in a given event, then we can form a total of
$N_c(N_c-1)/2$ $cc$ pairs ($\approx N_c ^2 /2 $ for $N_c \gg 1$)  and $N_c^2 $ $c\bar{c}$ pairs. In contrast, in the
case of $ N_{T_{cc}}^{(Mol)}$ and $ N_{X(3872)}^{(Mol)}$ we have
yields of the same order for both  $D D^*$ and $D \bar{D}^*$
pairs~\cite{Hu:2021gdg}; the difference comes from $\omega $.

\begin{center}
\begin{table}[h!]
  \caption{The $T_{cc}$ yields in central $Pb-Pb$ collisions at          
    $\sqrt{s_{NN}} = 5.02$ TeV at the LHC using the  coalescence model,  
    Eq.~(\ref{TccCoal}), for compact tetraquark ($4q$) and for molecular
    ($Mol$) configurations. } 
  \vskip3.0mm
\label{Tab2}
 \begin{tabular}{c | c c  }
\hline
\hline
State      & $N^{(4q)}(\tau_C)$     & $N^{ (Mol)}(\tau_H) $  \\   
\hline
$T_{cc}^+$ & $ 8.40 \times 10^{-5}$ & $ 4.10 \times 10^{-2}$
\\  
$X(3872)$  & $ 1.81 \times 10^{-4}$ & $ 7.50 \times 10^{-2}$
\\
\hline
\hline
\end{tabular}
\end{table}
\end{center}

As discussed in previous Section, to estimate the $T_{cc}$ production  
and absorption contributions we are using  the form factors and 
couplings calculated with  QCDSR, which are more appropriate to 
multiquark systems in a compact configuration. Indeed, in the three-point
correlation function  in the QCDSR calculation all the quark fields in the
current  are defined at the same space-time point (we refer the reader to
Ref.~\cite{Abreu:2021jwm} for a more detailed discussion).

\subsection{The $T_{cc}$  and $X(3872)$ abundances} 

Now we study the time evolution of the $T_{cc}$ abundance by solving     
Eq.~(\ref{rateeq}), with initial conditions computed with the coalescence
model and given in Table~\ref{Tab2}.
\begin{figure}[!ht]
\includegraphics[{width=8.0cm}]{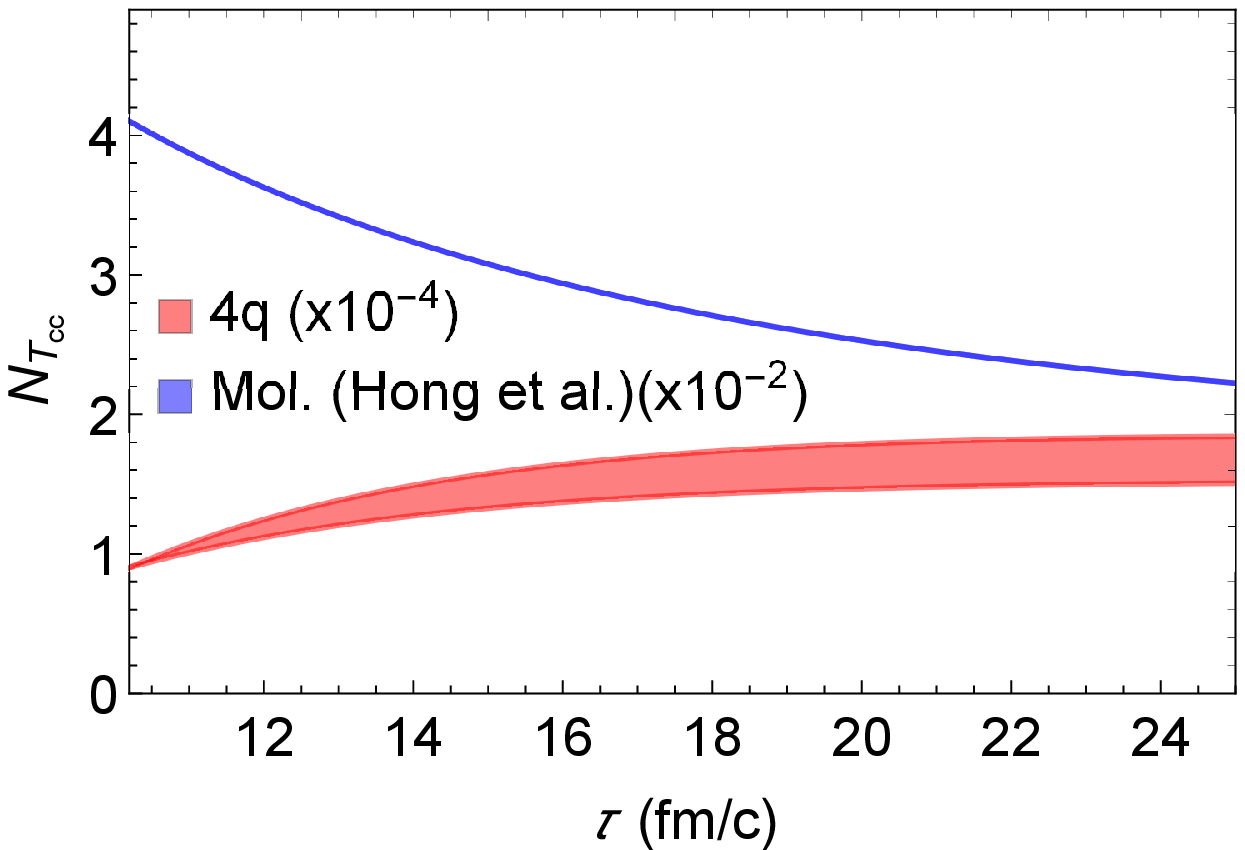}\\
\includegraphics[{width=8.0cm}]{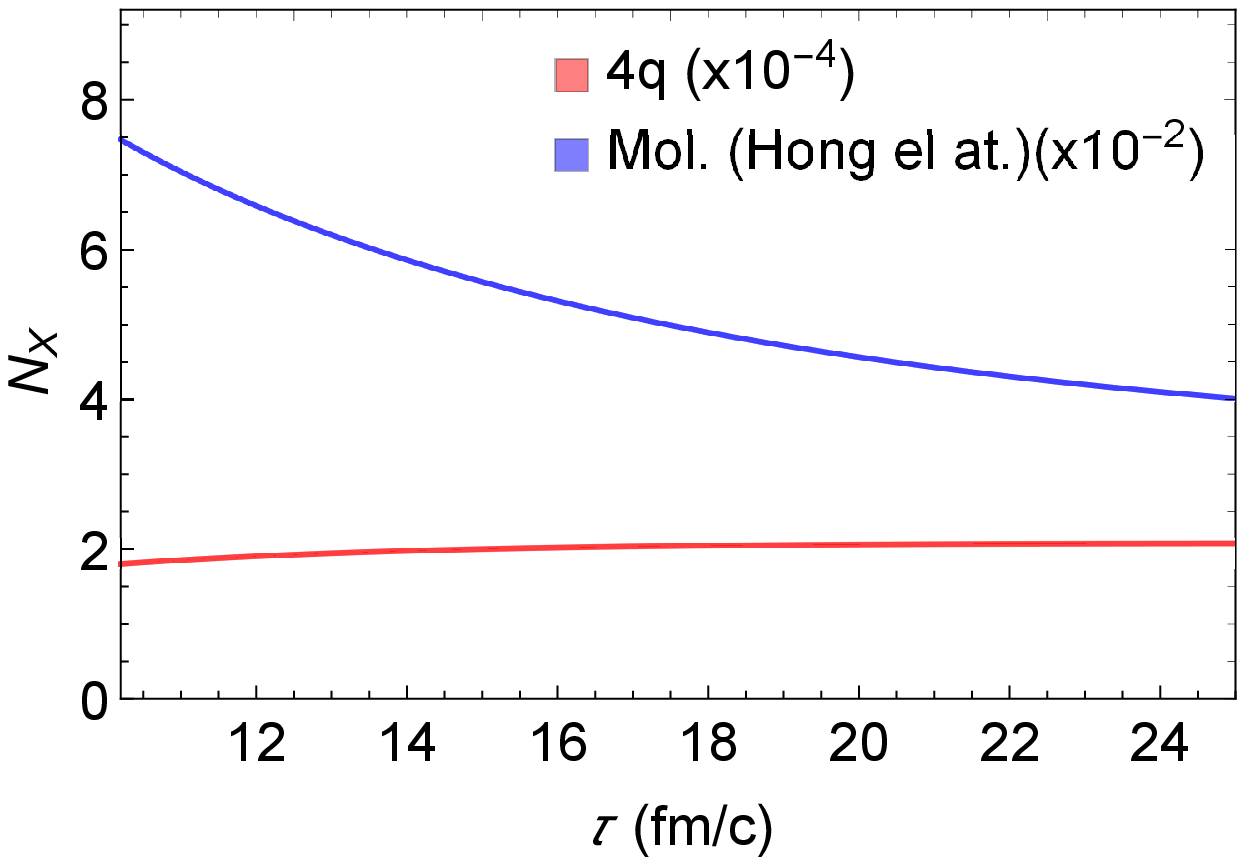}                  
\caption{a) Time evolution of the $T_{cc}$ abundance as a function
of the proper time in central $Pb-Pb$ collisions at $\sqrt{s_{NN}} = 5.02$
TeV, with initial conditions given by the  coalescence model. The tetraquark
and molecular abundances have been obtained from the present approach
(red curve), and from that described in Ref.~\cite{Hong:2018mpk} (blue curve). 
b) Same as a) but for the $X(3872)$ state.}
\label{TimeEvolTcc}
\end{figure}
In the  tetraquark configuration the $T_{cc}$  is a compact object and
its coupling constants and form factors can be computed with QCDSR, as
discussed in detail in \cite{Abreu:2021jwm}.  In the
molecular configuration, it is a very weakly bound state and its interaction
cross sections can be computed with the model proposed in
Ref.~\cite{Hong:2018mpk}. In this model, the authors make use of the         
``quasi-free'' approximation. The $D$ and the $D^*$ which form the $T_{cc}$,
are treated as approximately free particles which can interact with
the pions and $\rho$'s of the environment. The composite system is so weakly
bound that any of these interactions is able to destroy the $T_{cc}$.

In Fig.~\ref{TimeEvolTcc}a  we show the time evolution of the $T_{cc}$
abundance as a function of the proper time. In the  tetraquark curve,
the band represents the uncertainties coming from the 
QCDSR calculations of the absorption and production cross sections
\cite{Abreu:2021jwm}. In the figure we observe a strong sensitivity to the    
initial  yields. This can be understood looking at the two terms on the right
side of the rate equation~(\ref{rateeq}): the first, positive, is the ``gain''
term, whereas the second, negative, is the ``loss'' term, which depends on
$N_{T_{cc}}$. When $N_{T_{cc}}$ is initially very small (as it is for 
tetraquarks) the second term is very small, the first term dominates, the
derivative $d N_{T_{cc}} / d \tau$ is positive and the abundance of $T_{cc}$
increases. When $N_{T_{cc}}$ is initially large, the second term is bigger than
the first, derivative
$d N_{T_{cc}} / d \tau$ is negative and the abundance of $T_{cc}$ decreases. 
We conclude that, for  tetraquarks, $ N_{T_{cc}}^{(4q)}$ increases by a
factor of $\simeq 2$ during the hadron gas phase. For molecules the
absorption and regeneration terms yield similar contributions, with
predominance of the former and   $ N_{T_{cc}}^{ (Mol)} $ decreases. 
Comparing the final yields $N^{(4q)}_{T_{cc}} (\tau_F)$ and
$ N^{(Mol)}_{T_{cc}} (\tau_F)$, shown in Fig.~\ref{TimeEvolTcc}a,            
we find that for the molecular configuration the number of $T_{cc}$'s at the 
end of the hadron gas phase is two orders of magnitude larger! 
The difference in multiplicities decreases during the hadron gas phase but
it is still large at the end of the collision.

For the sake of comparison in Fig.~\ref{TimeEvolTcc}b we show
a plot similar to the one in Fig.~\ref{TimeEvolTcc}a, but for the $X(3872)$
state.
We remark that the time evolution of $ N_{X(3872)}  $ has already been analyzed
in Refs.~\cite{ChoLee1,XProd2} but in both cases the HIC environment chosen has
been the central $Au-Au$ collisions at $\sqrt{s_{NN}} = 200$ GeV at RHIC.
In  order to make a fair comparison between the $T_{cc}$ and $X(3872)$ yields,
we have redone the calculations of Refs. \cite{ChoLee1,XProd2} using the       
analogous reaction mechanisms for both states. As already mentioned above,
the  form factors in the vertices of the $T_{cc}$ reactions have been
calculated with QCD sum rules. 
Unfortunately the equivalent vertices for the $X(3872)$ are not available in
QCDSR. Thus, in this latter case, we have followed  \cite{ChoLee1,XProd2} and
used empirical monopole form factors taking the cutoff $\Lambda = 2.0$ GeV.

In the
present calculation of the tetraquark cross sections and time evolution, we
have ignored the terms with anomalous couplings, both in the $X(3872)$ and
$T_{cc}$ interactions. This is because the required
coupling $T_{cc} D^* D^*$ is not yet available and its computation is beyond the scope of the present work.
We are using the model of~\cite{Hong:2018mpk} for the molecules and
in this model the coupling $T_{cc} D^* D^*$ does not exist. Therefore
molecules are not affected by the lack of the anomalous couplings, only
tetraquarks. The similarities in mass and quantum numbers between $X(3872)$ and $T_{cc}$ suggest that the inclusion of the anomalous couplings interactions would reduce the multiplicity of tetraquark $T_{cc}$'s as it did for the $X(3872)$ (see Ref. [34]). The procedure adopted here allows for a
fair comparison between $X(3872)$ and $T_{cc}$ tetraquarks. However, because
of the approximations involved, our results should be regarded as upper limits for the multiplicities.

To summarize: 
using QCDSR and the quasi-free model we are able to perform a fair comparison
between the tetraquark (with both tetraquark initial
conditions and cross sections) and molecular
(with both molecular initial conditions and cross sections) approaches.  
We observe that if the $T_{cc}$ is a molecule,  
it will be produced more abundantly than a tetraquark 
and its multiplicity will decrease with time.
In contrast, tetraquarks would be produced much less abundantly and their
multiplicity would grow  with time. The difference in
multiplicities decreases during the hadron gas phase but
it is still large at the end of the collision.  We conclude that molecules
will be much more abundant (by a factor 100) than tetraquarks. This is in
qualitative agreement with the results found in \cite{Hong:2018mpk}.
The results for the $X(3872)$  are quite similar and indicate
that, even after going through the hadron gas phase, molecules remain
much more abundant than tetraquarks.


\section{System size dependence} 

Now we focus on the dependence of our results with the system size,
represented here by the density of charged particles measured at midrapidity
$\mathcal{N} = \left[ d N_{ch} / d \eta (|\eta| < 0.5)\right]^{1/3}$.
In our calculation we need to take into account the dependence of all the
relevant quantities with the charged particle multiplicity. As it can be seen from
Eq. ~(\ref{TccCoal}), the initial number of $T_{cc}$ tetraquarks depends on
$N_c$ and on the volume $V_C$. The initial number of $T_{cc}$ molecules depends
on $N_D$ and on the volume $V_H$.  All these quantities depend on the system size,
$\mathcal{N}$. The advantage of expressing the multiplicities in terms
of $\mathcal{N}$, instead of $\tau$, is that the former is a measurable quantity. 
In what follows we explain how we  can incorporate this
dependence in our formalism.

\subsection{Kinetic freeze-out time and temperaure}

As discussed
in Ref.~\cite{LeRoux:2021adw}, $\mathcal{N}$ may be empirically related to the
kinetic freeze-out temperature via the expression
\begin{equation}
T_F  = {T_{F0}} \, e^{- b \, \mathcal{N}},
\label{chiafit}
\end{equation}
where $T_{F0}$ and $b$ are constants chosen in order to fit the results of the
blastwave model analysis of the data performed by the ALICE Collaboration in 
\cite{alice13}. The freeze-out temperature depends on the system size.  This is
not surprising and has been realized long ago~\cite{hama92}. 
The values used here are the same of~\cite{LeRoux:2021adw}: 
$T_{F0} = 132.5$ MeV and $ b = 0.02$. Assuming that the hadron gas undergoes a 
Bjorken-like cooling, the freeze-out time, $\tau_F$, can be related to the
freeze-out temperature, $T_F$,  through the expression,
\begin{equation}
\tau_F = \tau_H  \left( \frac{T_H}{T_F} \right)^3.
\label{bjorf} 
\end{equation}
Inserting  Eq.~(\ref{chiafit}) into the above relation, we find:
\begin{equation}
\tau_F =  \tau_H \left( \frac{T_H}{T_{F0}} \right)^3 e^{3 b \mathcal{N}}.
\label{relf}
\end{equation}
With the above expression, from the observable quantity $\mathcal{N}$ we can
infer the duration of the hadronic phase. As it can be seen, larger systems
produce more particles, a larger $\mathcal{N}$ and live longer. 
Hence, the use of Eq.~(\ref{relf}) in the solutions of Eq.~(\ref{rateeq}) allows
to calculate $N_{T_{cc}}$  and $N_{X(3872)}$ as a function  $\mathcal{N}$.

\subsection{The volume}

In Ref.~\cite{Vovchenko:2019kes}  the authors used the Statistical Hadronization
Model (SHM) to perform an
extensive fit of several hadron yields measured by the ALICE collaboration in
different centrality bins, at different energies and in $p-p$, $p-Pb$ and
$Pb-Pb$ collisions. In their Fig. 4, they present the relation between the
volume per rapidity slice, $dV/dy$, and the central multiplicity density,
$dN_{ch}/d\eta$, which they parametrize as
$$
\frac{d V}{d y} = 2.4 \, \frac{d N_{ch}}{d \eta} (|\eta|<0.5)
= 2.4 \, \mathcal{N}^3
$$ 
Integrating this equation over the appropriate rapidity interval, we find
$V = const \, \mathcal{N}^3$. The constant can be determined by imposing that
$V = 5380 $ fm${}^3$ when $\mathcal{N} \approx 12.43 $ (and
$\left[ d N_{ch} / d \eta (\eta < 0.5)\right] = 1908$~\cite{Niemi:2015voa}).
We finally obtain:
\begin{equation}
V = 2.82 \, \mathcal{N}^3
\label{relV}
\end{equation}
In the context of the SHM,  $V$ is the chemical freeze-out volume. Here
we will follow  Ref.~\cite{ExHIC:2017smd} and assume that $V = V_H = V_C$.

\subsection{The number of charm quarks}

As far as we can tell, there is no experimentally established connection between
$N_c$ and $d N_{ch} / d \eta (\eta < 0.5)$. We will make use of the only work
where this relation was studied. In ~\cite{ALICE:2015ikl} the ALICE collaboration
measured the production of charm mesons in $pp$ collisions at $\sqrt{s} = 7$ TeV.
In the Fig. 2 of that paper there is plot of the differential distribution of
$D$ mesons as a function of $d N_{ch} / d \eta $. The experimental points can be
parametrized by a power law: 
$$
\frac{d^2N_D}{d y \, d p_T} / \langle \frac{d^2N}{d y \, d p_T} \rangle 
= {\alpha}' \, \left(\frac{d N_{ch}}{d \eta } /
\langle \frac{d N_{ch}}{d \eta } \rangle \right)^{\beta}  
$$
where the quantities in brackets are average values.
Fitting the experimental points we find that $\beta = 1.6$.
Integrating over the appropriate interval of 
rapidity and transverse momentum and rearranging the constants we arrive at:
\be
N_D = {\alpha}^{''} \, \left(\frac{d N_{ch}}{d \eta } \right)^{\beta}
    = {\alpha}^{''}  \, \left(\mathcal{N}^3\right)^{\beta} 
\label{relND}
\ee
We further assume
that the number of charm quarks and the number of $ D $ mesons are proportional: 
\be
N_c = const.  N_D = \alpha \, \left(\mathcal{N}^3\right)^{\beta} 
\label{propcD}
\ee
The constant $\alpha$ can be determined by  using the numbers shown in
Table~\ref{param}. 
In particular, we must have $N_c = 14$ for $\mathcal{N} = 12.43$.  We finally
arrive at:
\begin{equation}
N_c =  7.9 \times 10^{-5} \, \mathcal{N}^{4.8}.
\label{relNc}
\end{equation}

\begin{figure}[!ht]
\begin{center}
\includegraphics[{width=8.0cm}]{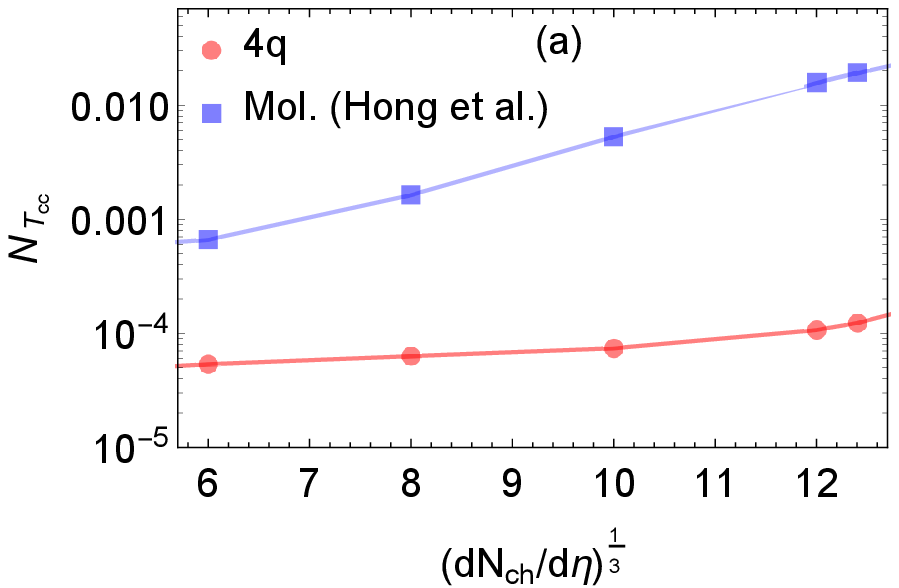}\\
\vskip0.2cm
\includegraphics[{width=8.0cm}]{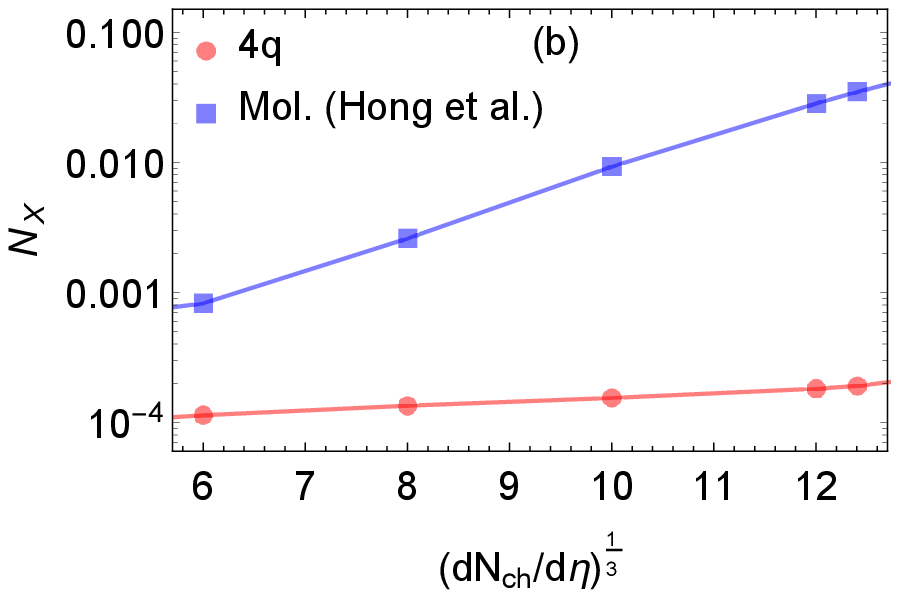}\\
\vskip0.2cm 
\includegraphics[{width=8.0cm}]{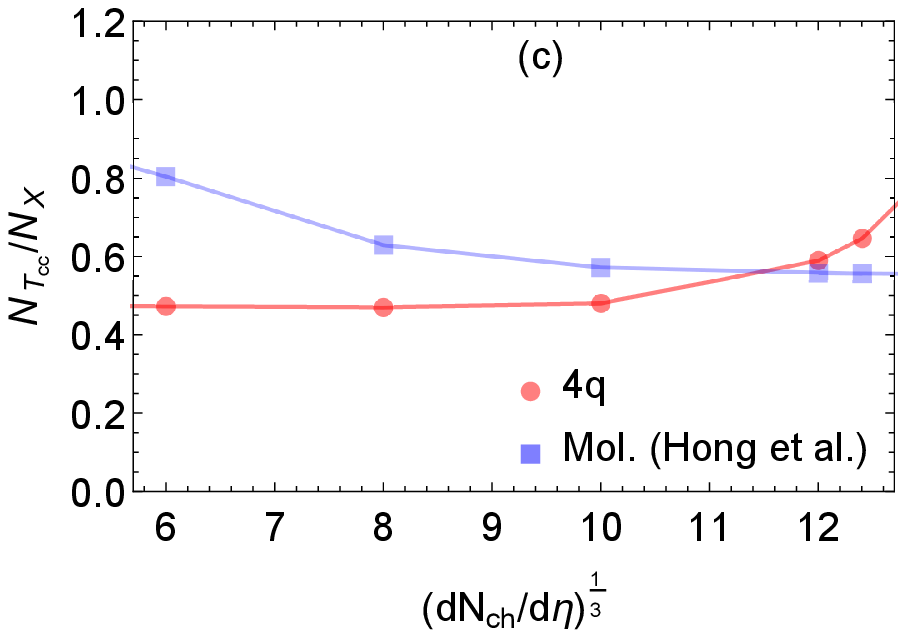}
\end{center}
\caption{a) The $T_{cc}$ abundance as a function
of  $\left[ d N / d \eta (\eta=0)\right]^{1/3}$, with initial conditions
given by the coalescence model.  b) Same as a) but for the $X(3872)$ state.
c) The ratio between the $T_{cc}$ and $X(3872)$ abundances as a function
of  $\left[ d N / d \eta (\eta=0)\right]^{1/3}$, with initial conditions
given by the coalescence model.}
\label{radata}
\end{figure}

Having established relations between the relevant quantities and
$\mathcal{N}$, we
proceed as follows. We first choose the system size parameter, $\mathcal{N}$.
Then, substituting (\ref{relV}), (\ref{relND})  and (\ref{relNc}) into
(\ref{TccCoal}) we obtain the $\mathcal{N}$ dependent initial conditions. In the
coalescence model, the number of composite particles depends on the volume. This
dependence is very different if the number of constituents is two (molecules) or
four (tetraquarks). Indeed, from (\ref{TccCoal}) we have $ N^{mol} \propto 1/V$
and $ N^{4q} \propto 1/V^3$. The number of available charm quarks or $D$ mesons
also depends on the volume of the system. However, as we can see  from
(\ref{propcD}), this dependence will be the same for molecules and tetraquarks.
With the above mentioned substitutions we arrive at:
\be
N^{4q} \propto \mathcal{N}^{0.6}  \hskip1cm \mbox{and} \hskip1cm
N^{mol} \propto \mathcal{N}^{6.6} 
\label{ndep}
\ee
This implies that, as we go to smaller systems (smaller $\mathcal{N}$), the
difference in the predictions tend to quickly disappear. This trend will not
be changed  during the evolution of the hadron gas, which, as shown in
Fig.~\ref{TimeEvolTcc}, only produces mild changes in the multiplicities.

Using the initial
conditions, we integrate the evolution equation (\ref{rateeq}) and stop 
at the kinetic freeze-out time, $\tau_F$, given by  (\ref{relf}) (and thus
carrying another $\mathcal{N}$ dependence).

Following this procedure we can replot Fig.~\ref{TimeEvolTcc} in terms of
the variable $\mathcal{N}$. We show the new plots in Fig.~\ref{radata}a and
Fig.~\ref{radata}b for the $T_{cc}^+$ and $X(3872)$ respectively. 
We observe that, not surprisingly,  both tetraquark and molecule multiplicities
increase as the system size grows. However the number of molecules increases much
more. This suggests that collisions with heavier ions are really more useful to
discriminate between the two configurations.

The ratio between the $T_{cc}^+$ and $X(3872)$ abundances, shown in
Fig.~\ref{radata}c, has a very  interesting behavior. In order to 
form a $X(3872)$ we need to produce one $c - \bar{c}$ pair. On the other hand, 
the formation of one $T_{cc}^+$ requires the production of two
$c - \bar{c}$ pairs. Therefore, in collisions of small systems one would
expect  the ratio
$T_{cc}^+ / X(3872)$ to be roughly one half.  Moving to larger systems, due
to the interactions with the medium, this difference tends to decrease and
the ratio tends to grow. This is what we see
in tetraquark curve in Fig.~\ref{radata}c. For tetraquarks we can make
predictions of the final yields based almost only on the initial conditions.
This is so because, as one can see in  Fig.~\ref{TimeEvolTcc}, the tetraquark
abundances almost do not change during the evolution of the hadron gas. 
In the molecular approach the initial ratio is strongly affected by the powers
of the binding energies, which are different for $T_{cc}^+$ and $X(3872)$.
Moreover, in
this case, the evolution plays a more important role, depending on the details
of the interactions which are different for $T_{cc}^+$ and $X(3872)$. The
outcome of all these dependences is the falling curve in Fig.~\ref{radata}c.
This result suggests that the behavior of the $T_{cc}^+ / X(3872)$ ratio might
be useful to discriminate between tetraquarks and molecules. The subject
deserves further studies.

These results are predictions  that can be tested experimentally in the 
future.


\section{Conclusions}
\label{Conclusions}


In this work we have investigted the multiplicity evolution of the doubly  
charmed state $T_{cc}^+$ in a hot hadron gas produced in the late stage of  
heavy-ion collisions. Effective Lagrangians have been used to calculate the
thermal   cross sections of $T_{cc}^+$ production in reactions     
such as $T_{cc}^+ \pi,  T_{cc}^+ \rho \rightarrow D^{(*)} D^{(*)} $  and its
absorption in the corresponding inverse ones.  We have found that the magnitude 
of the thermally averaged cross sections for the dissociation and production
reactions differ in some cases by factors of some orders of magnitude. 

With the thermal  cross sections as input, we have solved the rate 
equation to determine the time evolution of the  $T_{cc}^+$ multiplicity,   
considering different internal structures in the context of the coalescence
model:   
$T_{cc}^+$ as a $S$-wave tetraquark produced via quark coalescence mechanism   
from the QGP phase at the critical temperature; and as a $S$-wave weakly bound
hadronic molecule from the coalescence of mesons $D D^*$ formed at the end of
the mixed phase.
The results have suggested that when the initial conditions from four-quark   
coalescence model are employed,  $ N_{T_{cc}}$ increases by a factor about 2 
at freeze-out. However, it is still two order of magnitude smaller than the
final yield of molecules formed from hadron coalescence.
Therefore, we believe that the measurement of the $T_{cc}^+$ abundance in HICs  
might help in  discriminating one structure from the other.  
We emphasize that we presented for the first time a fair comparison between
a ``pure'' molecule evolution and a ``pure'' tetraquark evolution through the
hadronic medium in the context of the Effective Lagrangian approach. We also
used here the connection between the  time evolution of the hadronic fireball
and the measured charged particle rapidity density (measured at midrapidities). 
Both for molecules and tetraquarks the multiplicities increase with
$\left[ d N / d \eta (\eta=0)\right]$.

Using the phenomenological relations involving $\mathcal{N}$, we were able
arrive at the predictions shown in Fig.~\ref{radata},  which can be compared
to data, when they will be available. Our results are encouraging. They suggest
that it is indeed possible to use heavy ion collisions to determine the internal
structure of the new heavy exotic states. We plan in the near future to carry out
these calculations in a more rigorous way, improving, among other things,
our techniques to estimate the relevant volumes.

\begin{acknowledgements}


  The authors would like to thank the Brazilian funding agencies for their
  financial support: CNPq (LMA: contracts 309950/2020-1 and 400546/2016-7),
  FAPESB (LMA: contract INT0007/2016) and INCT-FNA.

\end{acknowledgements}


\end{document}